\documentclass[epj]{webofc}
\usepackage[utf8]{inputenc}
\usepackage[varg]{txfonts}   
\usepackage{booktabs}
\usepackage{xcolor}
\definecolor{darkred}{rgb}{0.4,0.0,0.0}
\definecolor{darkgreen}{rgb}{0.0,0.4,0.0}
\definecolor{darkblue}{rgb}{0.0,0.0,0.4}
\usepackage[bookmarks,linktocpage,colorlinks,
    linkcolor = darkred,
    urlcolor  = darkblue,
    citecolor = darkgreen]{hyperref}
\wocname{EPJ Web of Conferences}
\woctitle{Lattice2017}
\begin{document}
%
\selectlanguage{english}
\title{%
Baryon magnetic moments: Symmetries and relations}
\author{%
\firstname{Assumpta} 
\lastname{Parre\~no}
\inst{1}
\and
\firstname{Martin J.} 
\lastname{Savage}
\inst{2}
\and
\firstname{Brian C.}  
\lastname{Tiburzi}
\inst{3,4}
\fnsep
\thanks{Speaker, \email{btiburzi@ccny.cuny.edu}}
\and
\firstname{Jonas} 
\lastname{Wilhelm}
\inst{5,6} 
\and
\firstname{Emmanuel} 
\lastname{Chang} 
\inst{2}
\and
\firstname{William} 
\lastname{Detmold} 
\inst{7}
\and
\firstname{Kostas} 
\lastname{Orginos} 
\inst{8,9}
}
\institute{%
Department of Quantum Physics and Astrophysics, 
and Institute of Cosmos Sciences, University of Barcelona, E-08028, Spain
\and
Institute for Nuclear Theory, University of Washington, Seattle, WA 98195, USA
\and
Department of Physics, The City College of New York, New York, NY 10031, USA
\and
Graduate School and University Center, The City University of New York, New York,
NY 10016, USA
\and
Justus-Liebig-Universit\"at Gie\ss en, Ludwigstra\ss e 23, Gie\ss en 35390, Germany
\and
Department of Physics, University of Washington, Seattle, WA 98195, USA
\and
Center for Theoretical Physics, Massachusetts Institute of Technology, Cambridge,
MA 02139, USA
\and
Department of Physics, College of William and Mary, Williamsburg, VA 23187, USA
\and
Jefferson Laboratory, 12000 Jefferson Avenue, Newport News, VA 23606, USA
}
\abstract{%
 Magnetic moments of the octet baryons are computed using lattice QCD in background magnetic fields, 
 including the first treatment of the magnetically coupled $\Sigma^0$-$\Lambda$ system. 
 Although the computations are performed for relatively large values of the up and down quark masses, 
 we gain new insight into the symmetries and relations between magnetic moments by working at a three-flavor mass-symmetric point. 
 While the spin-flavor symmetry in the large $N_c$ limit of QCD is shared by the na\"ive constituent quark model, 
 we find instances where quark model predictions are considerably favored over those emerging in the large $N_c$ limit. 
 We suggest further calculations that would shed light on the curious patterns of baryon magnetic moments.
}
\maketitle

\section{Introduction}
\label{intro}

The electromagnetic properties of hadrons and nuclei provide physically intuitive information about their structure.
Relations between baryon magnetic moments, 
moreover, 
were historically crucial in exposing the approximate symmetries of QCD.  
Here, 
we argue that these magnetic moments continue to provide
(increasingly subtle) 
clues about the structure of baryons.

In the past few years, 
the 
$\texttt{NPLQCD}$ 
collaboration has undertaken the first lattice gauge theory computations of the magnetic properties of light nuclei. 
Highlights of these computations include:
determination of the magnetic moments and polarizabilities of light nuclei%
~\cite{Beane:2014ora,Chang:2015qxa};
study of the simplest nuclear reaction, 
$n + p \to d + \gamma$, 
through its dominant magnetic dipole transition amplitude%
~\cite{Beane:2015yha}; 
and, 
uncovering hints of unitary nucleon-nucleon interactions in large magnetic fields%
~\cite{Detmold:2015daa}. 
These computations were made possible by two crucial ingredients:
i).~the external field technique, 
for which these properties can be determined from two-point correlation functions rather than three-point functions;
and
ii).~rather large quark masses, 
for which adequate statistics can be accumulated to provide signals for light nuclei. 
Here, 
we describe an extra curricular study that emerged from this work, 
namely the determination of octet baryon magnetic moments%
~\cite{Parreno:2016fwu}. 
While such computations are not new, 
the setup of our calculation enabled investigations of the magnetic moments that had not previously been made. 
Most notably, 
we study strong interactions in an unphysical environment: 
that of exact
SU$(3)_F$
symmetry, 
in which the quark masses satisfy
$m_u = m_d = m_s \approx (m_s)_\text{physical}$. 
The addition of the quark electric charges preserves an 
SU$(2)$
subgroup of 
SU$(3)_F$, 
commonly called 
$U$-spin.

A summary of this investigation is presented, 
including a brief overview of the calculational details, 
Sec.~\ref{s:over}. 
We discuss the issue of magneton units for moments, 
Sec.~\ref{s:nBM}, 
the determination and estimation of Coleman-Glashow magnetic moments,
Sec.~\ref{s:CG}, 
and compare predictions of the na\"ive constituent quark model with those of the large $N_c$ limit of QCD, 
Sec.~\ref{s:QMvsNc}. 
We review our treatment of the coupled system of 
$\Sigma^0$ 
and 
$\Lambda$
baryons in Sec.~\ref{s:siglam}. 
Questions for future study are given in the concluding section, 
Sec.~\ref{s:end}.

\section{Overview of Calculation}
\label{s:over}

The calculation of octet baryon magnetic moments in
Ref.~\cite{Parreno:2016fwu}
utilizes primarily two ensembles of tadpole-improved clover fermions, 
details of which have been given previously%
~\cite{Beane:2012vq,Orginos:2015aya}. 
As mentioned, 
one ensemble is 
SU$(3)_F$
symmetric, 
for which the pion mass is 
$m_\pi \approx 800 \, \texttt{MeV}$.
The other ensemble maintains 
$m_s \approx (m_s)_\text{physical}$, 
but with two degenerate lighter quarks, 
corresponding to a pion mass of 
$m_\pi \approx 450 \, \texttt{MeV}$.

To study the magnetic properties of hadrons, 
classical 
U$(1)$
gauge links are post-multiplied to the color links.%
\footnote{
The magnetic field is not included in the generation of QCD gauge fields. 
For the SU$(3)_F$ ensemble, 
the computation of magnetic moments is exact, 
due to the condition 
$\texttt{Tr} \left( Q  \right)= q_u + q_d + q_s = 0$. 
With broken SU$(3)_F$ on the second ensemble, 
however, 
contributions from coupling the magnetic field to the quark sea are missing. 
The size of such quark-disconnected contributions to isoscalar quantities has been estimated to be quite small, 
see, 
e.g., Ref.~\cite{Abdel-Rehim:2013wlz},
and will be neglected throughout. 
} 
To add a uniform magnetic field in the $x_3$-direction, 
we choose
\begin{equation}
U_\mu (x)
=
e^{- i q x_2 B \, \delta_{\mu 1}}
e^{+ i q x_1 B N \, \delta_{\mu 2} \, \delta_{x_2, N-1}}
\in 
\text{U}(1)
,\end{equation}
where 
$q$ 
is the quark charge, and the magnetic field,
$B$, 
must satisfy the 't Hooft quantization condition%
~\cite{tHooft:1979rtg}, 
namely
$q B = 2 \pi \, n_\Phi / N^2$, 
to ensure that the flux through each elementary plaquette of the lattice is uniform.  
The number of lattice sites in each spatial direction is taken as 
$N$, 
and 
$n_\Phi$
is the flux quantum of the torus. 
Quark propagators are computed for the values
$n_\Phi = 0$, $+3$, $-6$, and $+12$, 
where the overall factor of 
$3$
is due to the fractional nature of the quark charges, 
and the multiplicative factors of
$-2$
allow one to recycle some down-quark propagators as up-quark propagators. 
With 
$48$ 
random source locations per configuration 
and a combined total of 
$\sim 1500$ 
configurations for the two ensembles, 
on the order of 
$400 \, \texttt{k}$
measurements are made. 
For the two ensembles considered, 
the spatial size is
$N = 32$, 
leading to magnetic fields that are expected to be of a reasonable perturbative size.

\begin{figure} 
  \centering
    \resizebox{\linewidth}{!}{
    \includegraphics[height=3cm,clip]{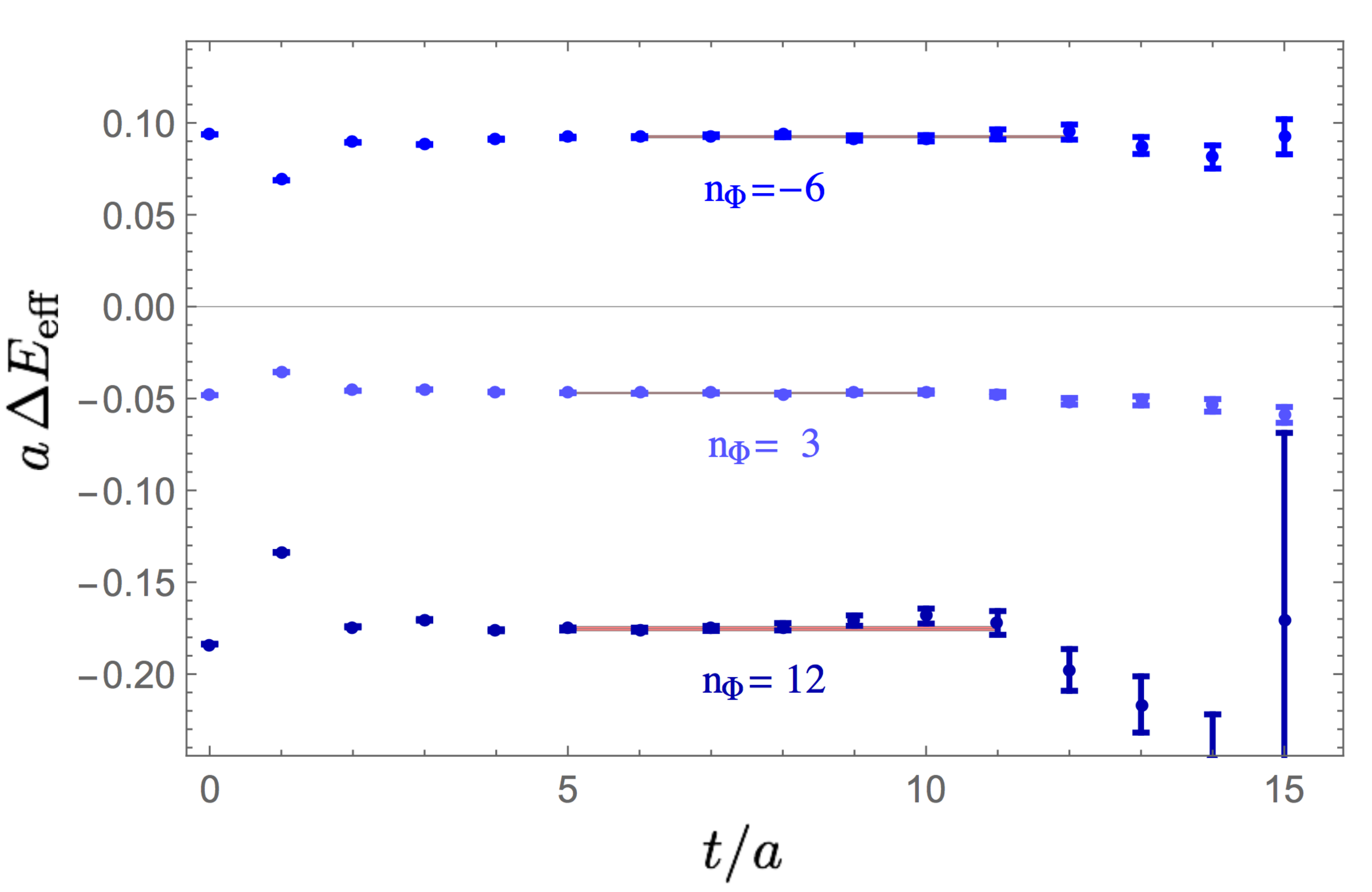}
  $\,$
  \includegraphics[height=3cm,clip]{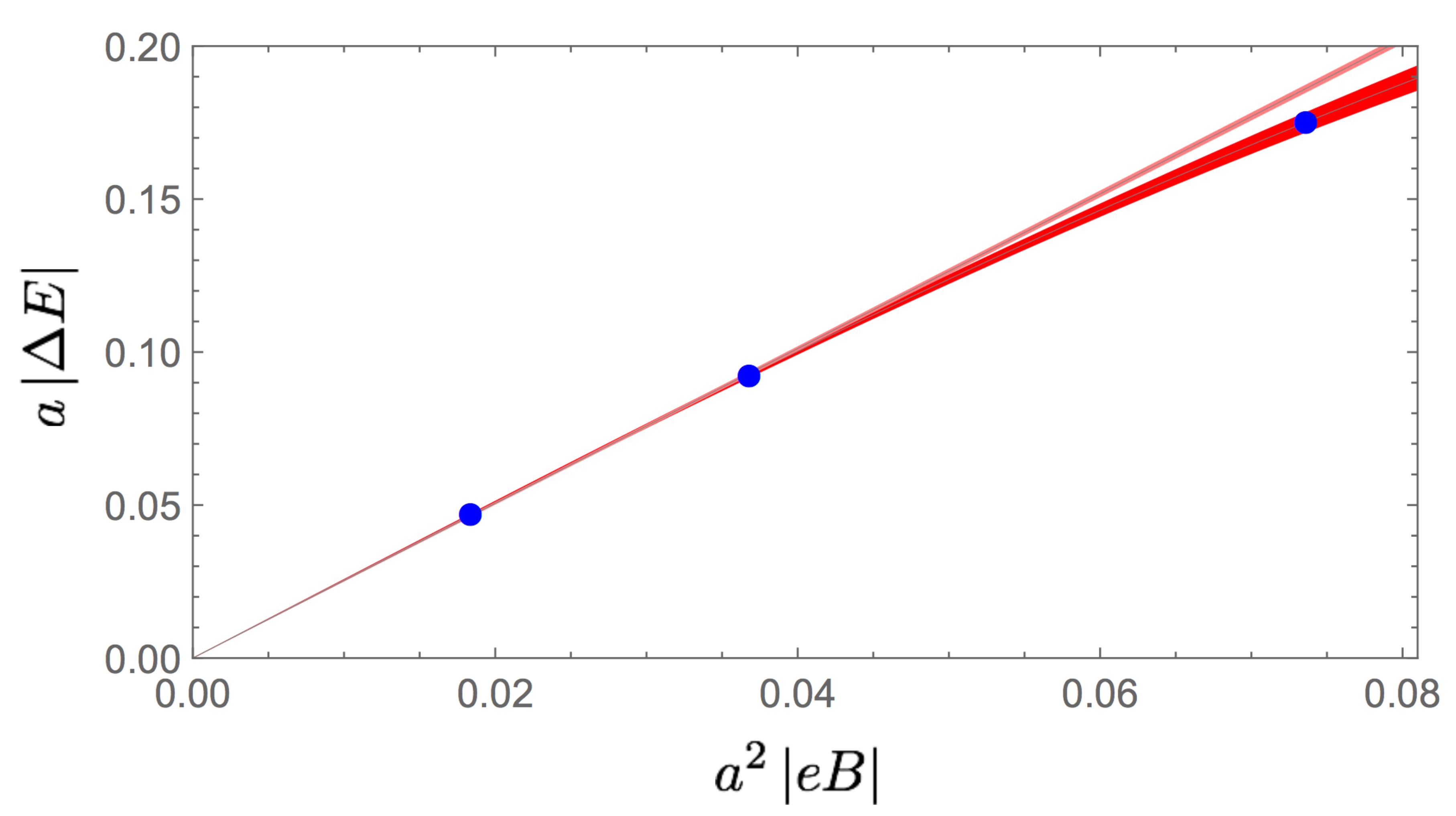}
  }
  \caption{Effective mass plots for the proton spin splittings in three magnetic field strengths appear on the left, 
  where bands show the results of fits to constant time dependence.  
  Magnetic field-strength dependence of the extracted proton spin splittings is shown on the right, 
  where bands show the results of linear and linear-plus-cubic fits, 
  with the largest field strength excluded in the former. 
  }
  \label{fig-1}
\end{figure}

The procedure to determine magnetic moments from external field calculations goes back to the early days of lattice QCD%
~\cite{Martinelli:1982cb,Bernard:1982yu}. 
Rather simply, 
one calculates the Zeeman effect for hadrons. 
This is accomplished by computing the spin-dependent hadron energies as a function of the magnetic field. 
Double ratios of correlation functions can be devised to cleanly isolate the splitting between spin states, 
e.g.
\begin{equation}
\frac{G^\uparrow (t, n_\Phi)}{G^\downarrow(t, n_\Phi) }
\Bigg/ 
\frac{G^\uparrow (t, 0)}{ G^\downarrow(t, 0) }
= 
Z \,
e^{ - \Delta E t}
+ 
\cdots
,\end{equation}
where we have exhibited saturation by the ground state in the long-time limit, 
and the spin splitting 
$\Delta E$
is given by 
$\Delta E = E^\uparrow - E^\downarrow = - 2 \mu B + \cdots$. 
Effective mass plots for the proton spin splittings in the various magnetic fields are shown in 
Fig.~\ref{fig-1}. 
Additionally shown is the behavior of these splittings as a function of the magnetic field.
Extraction of the proton magnetic moment is limited by the systematics of fitting the linear magnetic field-strength term, 
not the statistics of the present computation.

\section{Results}
\label{s:results}

Magnetic moments of the octet baryons with 
$I_3 \neq 0$
are determined using the procedure outlined above. 
In the following, 
we explore units for magnetic moments that suppress their pion-mass dependence, 
relations between the magnetic moments in the limit of 
SU$(3)_F$, 
as well as relations predicted in the constituent quark model and large $N_c$ limit. 
The two 
$I_3 = 0$
baryons, 
$\Sigma^0$
and
$\Lambda$,  
which undergo mixing in magnetic fields, 
are discussed separately.

\subsection{Natural Baryon Magnetons}
\label{s:nBM}

\begin{figure} 
  \centering
  \resizebox{\linewidth}{!}{
  \includegraphics[height=5cm,clip]{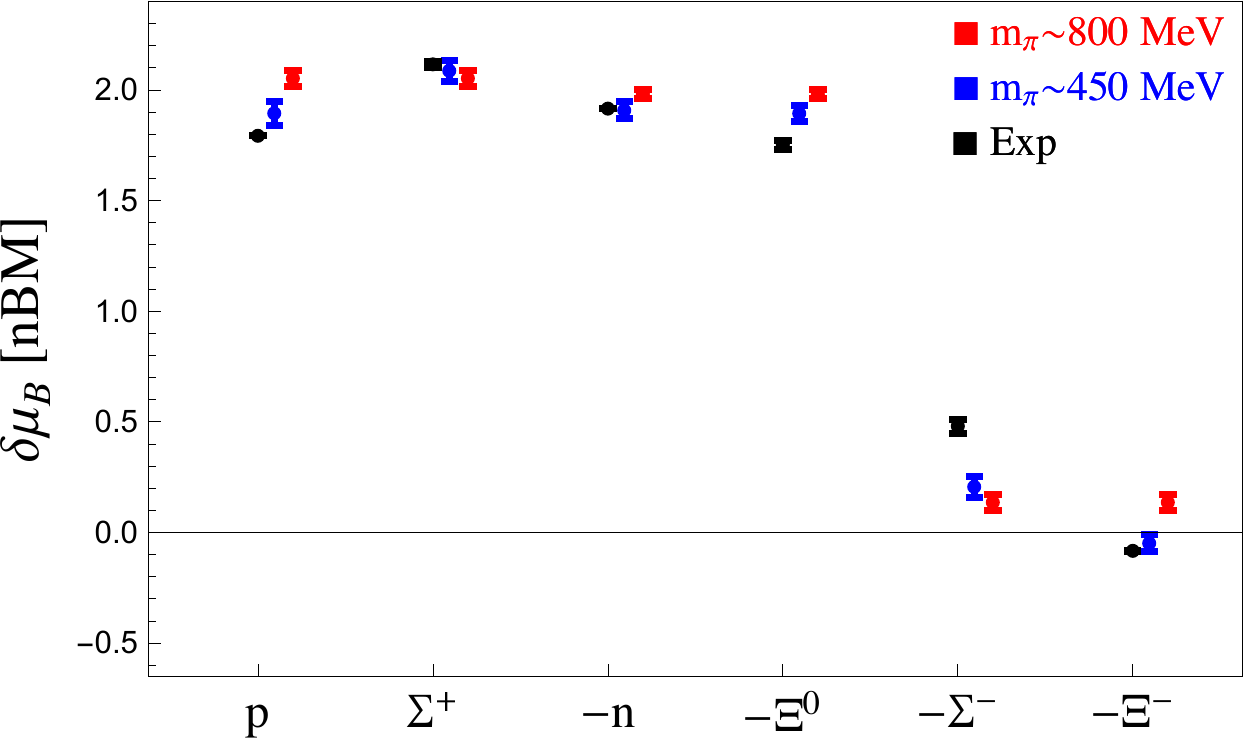}
  $\quad$
  \includegraphics[height=5cm,clip]{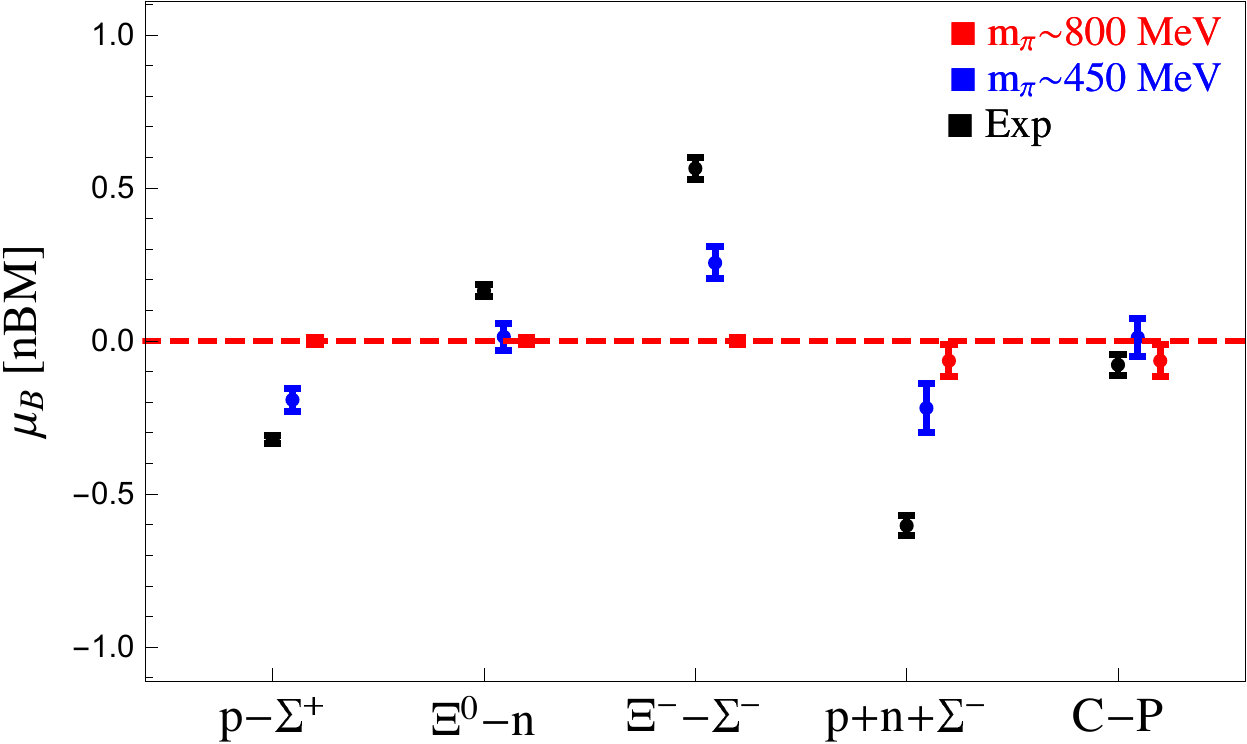}
   }
  \caption{Anomalous magnetic moments of the octet baryons in units of natural baryon magnetons appear on the left.
  We use the notation ``$-B$ '' to specify the negative of the magnetic moment, 
  $\delta \mu_{-B} \equiv - \delta \mu_B$, 
  so that a majority of the moments plotted are positive. 
  On the right appear linear combinations of magnetic moments that vanish in the limit of $U$-spin symmetry, 
  for example, 
  $p - \Sigma^+$
  corresponds to 
  $\mu_p - \mu_{\Sigma^+}$. 
  The notation ``C-P'' refers to half the sum of all moments, 
  $\mu_{\text{C--P}} = \frac{1}{2} \left[ \mu_p + \mu_{\Sigma^+} + \mu_n + \mu_{\Xi^0} + \mu_{\Sigma^-} + \mu_{\Xi^-} \right]$, 
  which is one of the relations of 
  Ref.~\cite{Caldi:1974ta}.
  }
  \label{fig-b}
\end{figure}

Using the procedure above, 
the magnetic moments are extracted in lattice magnetons, 
$\frac{1}{2} e a$, 
where 
$a$
is the lattice spacing, 
and are of order unity,  
suggestive of neither large volume nor discretization effects. 
Inspired by the observation that nuclear and nucleon magnetic moments are strikingly close to their physical values when expressed in terms of 
natural nuclear magneton units%
~\cite{Beane:2014ora}, 
we present magnetic moments in natural baryon magnetons. 
These units are defined by 
$\texttt{[nBM]} \equiv \frac{e}{2 M_B(m_\pi)}$, 
where 
$M_B(m_\pi)$
denotes the particular octet baryon mass determined at the corresponding pion mass. 
Such units incorporate the surprisingly linear pion-mass dependence observed for octet baryon masses%
~\cite{WalkerLoud:2008bp}. 
Results for anomalous magnetic moments, 
$\delta \mu_B  \, \texttt{[nBM]} = \mu_B \, \texttt{[nBM]} - Q_B$,
are displayed in 
Fig.~\ref{fig-b}
along with experimental values. 
A salient feature of the natural baryon magneton unit is that the Dirac part of the moment, 
which is short distance in nature, 
can readily be subtracted. 
The results shown exhibit curious behavior. 
Across all of these baryon magnetic moments, 
the majority of their pion-mass dependence is accounted for by employing 
$\texttt{[nBM]}$. 
These units appear to suppress the breaking of $U$-spin symmetry, 
which, 
among other things, 
predicts
$\mu_p = \mu_{\Sigma^+}$, 
$\mu_n = \mu_{\Xi^0}$, 
and
$\mu_{\Sigma^-} = \mu_{\Xi^-}$. 
The latter moments, 
$\mu_{\Sigma^-}$ 
and
$\mu_{\Xi^-}$, 
show little deviation from those of pointlike particles. 
In Fig.~\ref{fig-b}, 
we additionally show the linear combinations of magnetic moments that vanish in the limit of 
$U$-spin symmetry. 
These linear combinations are quite small; 
nonetheless, 
they confirm that 
SU$(3)_F$
breaking on the 
$m_\pi = 450 \, \texttt{MeV}$
ensemble lies intermediate to the 
$m_\pi = 800 \, \texttt{MeV}$ 
ensemble and experiment.

\subsection{Coleman-Glashow Moments}
\label{s:CG}

\begin{figure} 
  \centering
  \sidecaption
  \includegraphics[width=6.5cm,clip]{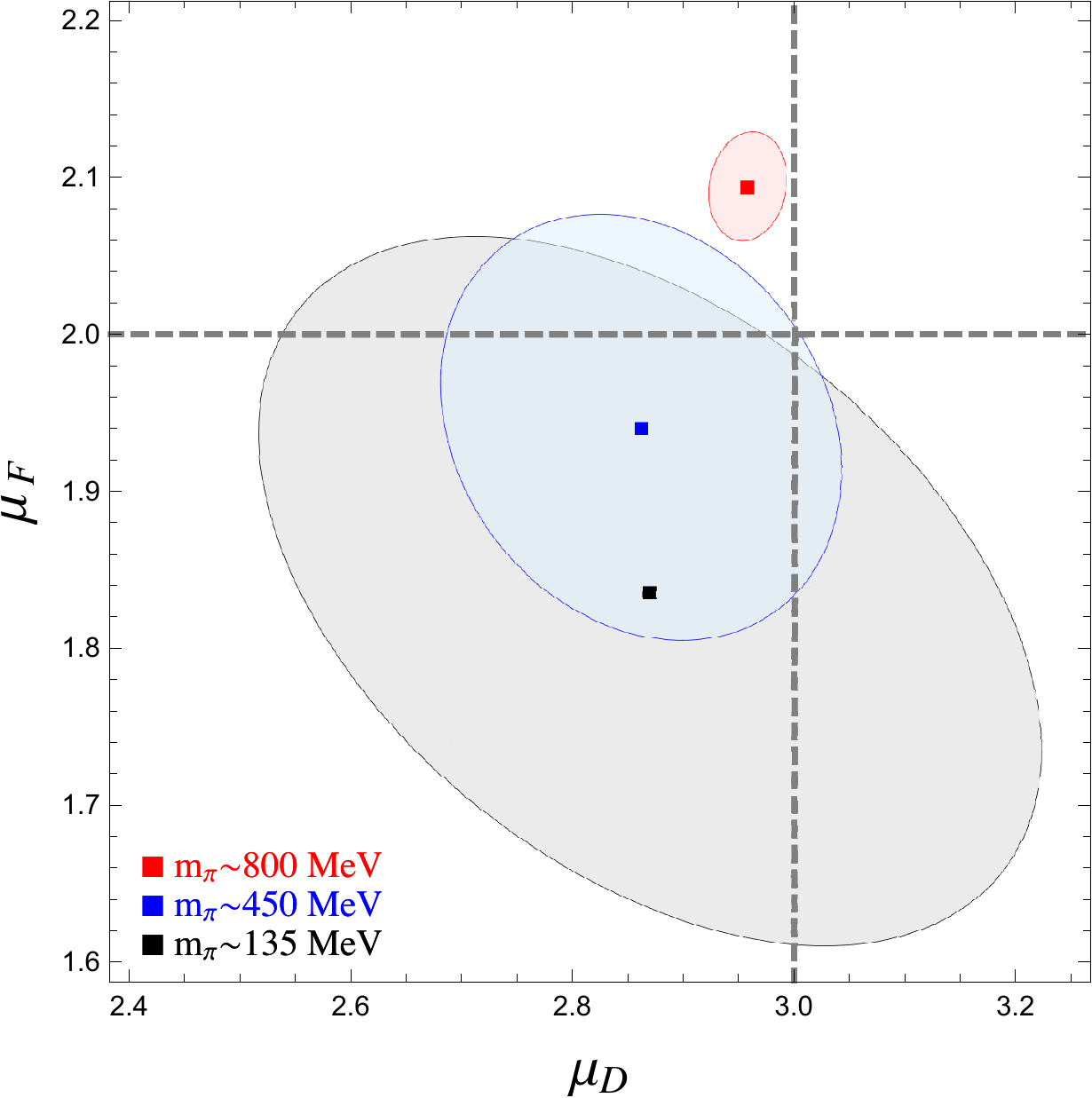}
  \caption{Values of the Coleman-Glashow magnetic moments, 
  $\mu_D$ and $\mu_F$, 
  determined from lattice QCD and experiment.
  Values are directly extracted from a combined fit to moments computed on the 
  $m_\pi = 800 \, \texttt{MeV}$ 
  ensemble, 
  due to $U$-spin symmetry. 
   Those determined from the 
   $m_\pi = 450 \, \texttt{MeV}$ 
   ensemble and experiment utilize only the nucleon magnetic moments, 
   and include an uncertainty due to the breaking of SU$(3)_F$ symmetry by the strange quark in the sea.
   This uncertainty is assessed as 
   $\Delta m_q  \, N_c^{-1}$, 
   where 
   $\Delta m_q$ 
   represents the expected size of SU$(3)_F$ breaking, 
   and the inverse factor of 
   $N_c = 3$ 
   reflects the closed fermion loop. 
   For experiment, 
   this uncertainty is taken to be
   $\Delta m_q \,  N_c^{-1} = 30\% \,  / \, 3 = 10\%$, 
   while for moments on the 
   $m_\pi = 450 \, \texttt{MeV}$ 
   ensemble, 
   we use 
   $\Delta m_q \, N_c^{-1} = 15 \% \, / \, 3 = 5 \%$, 
   which reflects the reduction in 
   SU$(3)_F$ 
   breaking exhibited in Fig.~\ref{fig-b}.}
  \label{fig-CG}
\end{figure}

In the limit of $U$-spin symmetry, 
the octet baryon magnetic moments are determined by only two numbers, 
$\mu_D$
and
$\mu_F$, 
which we call the Coleman-Glashow moments. 
These appear as coefficients in the effective Hamiltonian density%
~\cite{Coleman:1961jn}
\begin{equation}
\mathcal{H}
=
- 
\frac{e \, \vec{\sigma} \cdot \vec{B}}{2 M_B}
\left[
\mu_D \,
\texttt{Tr}
\left( \overline{B} \left\{ Q, B \right\} \right)
+ 
\mu_F \,
\texttt{Tr}
\left( \overline{B} \left[ Q, B \right] \right)
\right]
,\end{equation}
where 
$Q$
is the quark charge matrix, 
and
$B$ 
the octet of baryon fields, 
with 
$\overline{B}$
as its conjugate. 
Values of 
$\mu_D$
and
$\mu_F$
can be extracted on the 
$m_\pi = 800 \, \texttt{MeV}$
ensemble with high precision. 
Values can be determined from the 
$m_\pi = 450 \, \texttt{MeV}$
ensemble and experiment with additional assumptions. 
Determinations of these parameters are shown in 
Fig.~\ref{fig-CG}, 
and exhibit little pion-mass dependence.

\subsection{Quark Model vs. Large $N_c$}
\label{s:QMvsNc}

Investigation of the Coleman-Glashow moments suggests little deviation from 
$U$-spin symmetry, 
however, 
this is not the complete story. 
Results are
(not surprisingly)
consistent with a larger symmetry group, 
that of 
SU$(6)$
and the na\"ive consitituent quark model. 
Assuming non-interacting constituent quarks, 
one arrives at the values
$\mu_D = 3$
and
$\mu_F = 2$, 
where the former is simply counting the number of constituent quarks in each baryon. 
Such values, 
moreover, 
explain why the anomalous magnetic moments shown in 
Fig.~\ref{fig-b}
only take on values close to
$\delta \mu_B = 0$, 
and 
$\pm 2$.

\begin{figure}
  \centering
  \resizebox{\linewidth}{!}{
  \includegraphics[height=5cm]{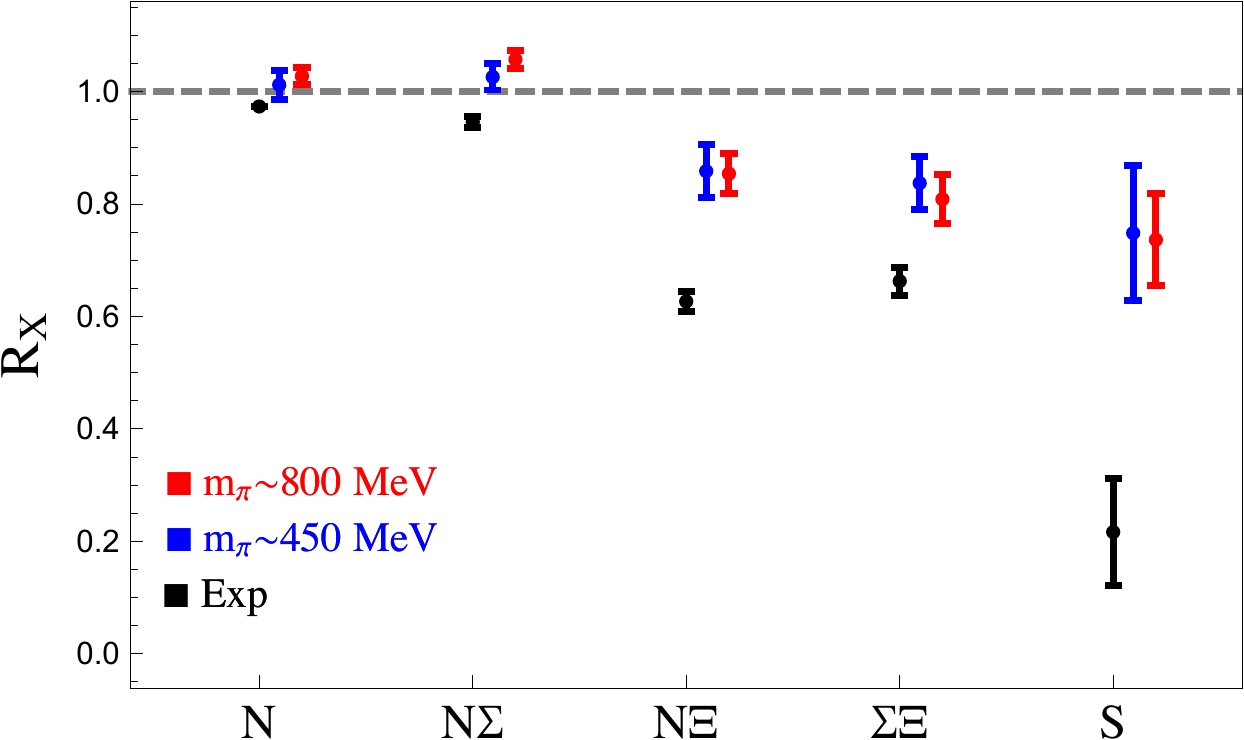}
  $\quad$
  \includegraphics[height=5cm]{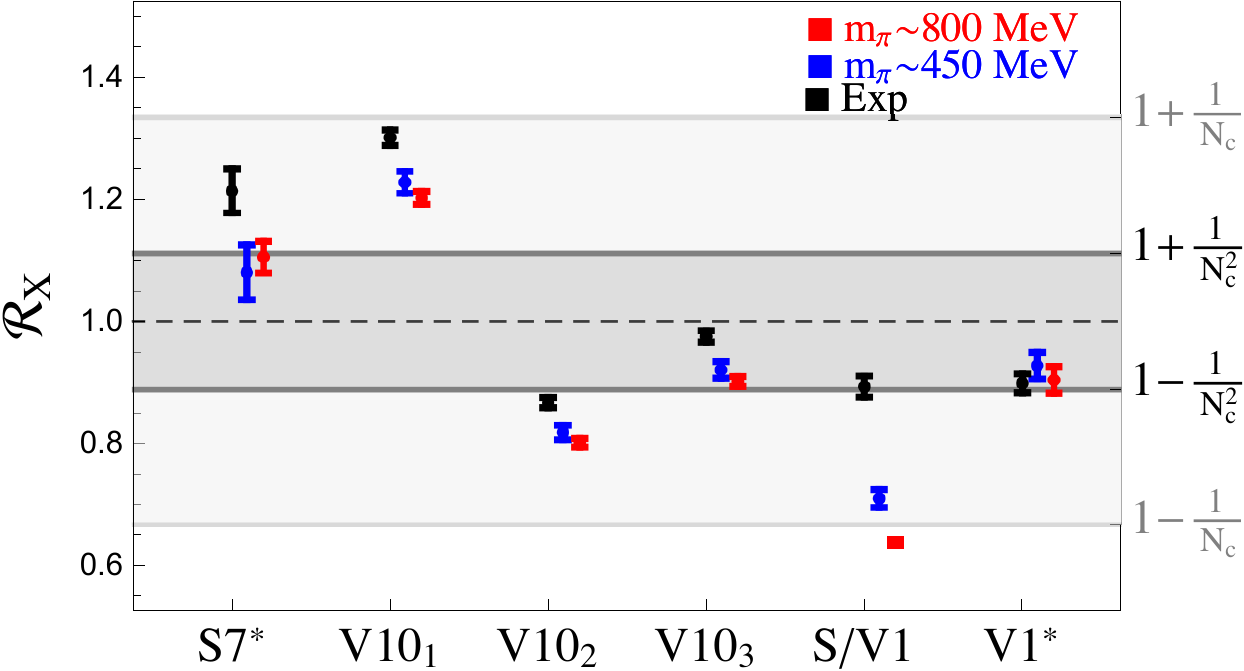}
  }
  \caption{Quark model (left) and large-$N_c$ relations (right) between magnetic moments. 
  All relations are given as ratios that are predicted to be unity.
  The quark model ratios are given in Eq.~\ref{eq:QM}, 
  while those emerging in the large $N_c$ limit are given in Eqs.~\ref{eq:Nc1} (starred) and \ref{eq:Nc2} (unstarred). 
  }
  \label{fig-QMvsNc}
\end{figure}

The lattice QCD results can be utilized to explore quark model relations as a function of the pion mass. 
In Fig.~\ref{fig-QMvsNc}, 
we show various ratios,
$R_X$, 
that are predicted to be unity within the na\"ive constituent quark model. 
Using 
$\Delta \mu$'s 
to denote the isovector differences of moments, 
these ratios are
\begin{equation}
R_N
= -\frac{2}{3} \frac{\mu_p}{\mu_n}
,\quad
R_{N \Sigma}
= 
\frac{5}{4} \frac{\Delta \mu_{\Sigma}}{\Delta \mu_N}
,\quad
R_{N \Xi}
=
5 \, \frac{\Delta \mu_\Xi}{\Delta \mu_N}
,\quad
R_{\Sigma \Xi} 
=
4 \, \frac{\Delta \mu_\Xi}{\Delta \mu_\Sigma}
,\quad 
\text{and}
\quad
R_{S}
=
- 4 \, \frac{\mu_{\Sigma^+} + 2 \mu_{\Sigma^-}}{\mu_{\Xi^0} + 2 \mu_{\Xi^-}}
\label{eq:QM}
.\end{equation}
Unity is very well satisfied for the first two ratios over the range of pion masses available. 
The remaining ratios show greater deviation, 
although the lattice QCD results generally show better agreement with the na\"ive quark model than experiment.  
One should note that the isovector moment ratios essentially compare two determinations of the (light) constituent quark mass, 
and that the ratio
$R_{N \Xi}$, 
for example, 
shows an 
\emph{environmental sensitivity} 
described in 
Ref.~\cite{Leinweber:1999nf}. 
The ratio
$R_S$ 
compares the \emph{in situ} determination of the strange constituent quark mass, 
which appears to be effectively different in 
$\Sigma$
and
$\Xi$
baryons.  
The pattern shown in the figure is perhaps suggestive of a systematic expansion scheme, 
and remains to be explored quantitatively.

From a field-theoretic perspective, 
the success of the non-relativistic quark model is argued to be understood due to the emergent spin-flavor symmetry of the 
large-$N_c$ limit%
~\cite{Dashen:1993jt,Dashen:1994qi}. 
Indeed the scaling of magnetic moment relations predicted in the large $N_c$ limit seems to explain why some of the quark model relations 
work better than others%
~\cite{Jenkins:1994md}, 
despite the restriction of nature to 
$N_c = 3$. 
The lattice calculations of 
Ref.~\cite{Parreno:2016fwu}, 
which are also restricted to 
$N_c = 3$,
offer further insight into large 
$N_c$
relations among baryon magnetic moments.
Results are shown in Fig.~\ref{fig-QMvsNc}; 
and, 
in the case of quark-mass independent relations, 
large $N_c$ 
scalings appear to be confirmed. 
These are predictions for the 
(starred) 
ratios
\begin{equation}
\mathcal{R}_{S7}
=
\frac{ 5( \mu_p + \mu_n) - (\mu_{\Xi^0} + \mu_{\Xi^-})}{4 ( \mu_{\Sigma^+} + \mu_{\Sigma^-})}
= 
1 + \mathcal{O}(N_c^{-1})
, \quad \text{and} \quad
\mathcal{R}_{V1}
=
\frac{\Delta \mu_N + 3 \Delta \mu_\Xi}{2 \Delta \mu_{\Sigma}}
= 
1 + \mathcal{O} (N_c^{-2})
\label{eq:Nc1}
.\end{equation}
By contrast,  
relations that depend on the level of 
SU$(3)_F$
breaking seem to exhibit a trend opposite large $N_c$ considerations. 
These are the (unstarred) ratios 
\begin{eqnarray}
\mathcal{R}_{V10_1}
&=&
\frac{\Delta \mu_N}{\Delta \mu_\Sigma}
=
1 + \mathcal{O} (N_c^{-1})
\overset{\text{SU}(3)_F}{=}
\begin{pmatrix}
1 + \mathcal{O} (N_c^{-1})
\\
1.25, \text{ CQM}
\end{pmatrix}
,\notag \\
\mathcal{R}_{V10_2}
&=&
\left( 1 - N_c^{-1} \right)
\frac{\Delta \mu_N}{\Delta \mu_\Sigma}
=
1 + \mathcal{O} ( \Delta m_q  N_c^{-1})
\overset{\text{SU}(3)_F}{=}
\begin{pmatrix}
1 + \mathcal{O} (  N_c^{-2})
\\
0.83, \text{ CQM}
\end{pmatrix}
,\notag \\
\mathcal{R}_{V10_3}
&=&
\left( 1 + N_c^{-1} \right)^{-1}
\frac{\Delta \mu_N}{\Delta \mu_\Sigma}
=
1 + \mathcal{O} ( \Delta m_q   N_c^{-1})
\overset{\text{SU}(3)_F}{=}
\begin{pmatrix}
1 + \mathcal{O} (  N_c^{-2})
\\
0.94, \text{ CQM}
\end{pmatrix}
,
\notag\\
\mathcal{R}_{S/V1}
&=&
\frac{\mu_p + \mu_n - 6 \left(N_c^{-1} - 2 N_c^{-2} \right) \Delta \mu_N}
{2 \left( \mu_{\Sigma^+} + \mu_{\Sigma^-} \right) -  \left( \mu_{\Xi^0} + \mu_{\Xi^-} \right)}
=
1 + \mathcal{O} ( \Delta m_q )+ \mathcal{O} ( \Delta m_q  N_c^{-1})
\overset{\text{SU}(3)_F}{=}
\begin{pmatrix}
1 + \mathcal{O} ( N_c^{-2})
\\
0.62, \text{ CQM}
\end{pmatrix}
,\notag \\
\label{eq:Nc2}
\end{eqnarray}
where we have also listed the values obtained in the 
SU$(3)_F$-symmetric constituent quark model (CQM). 
From the figure, 
notice that scaling of the ratios
$\mathcal{R}_{V10_2}$
and
$\mathcal{R}_{V10_3}$
does not seem to improve with 
SU$(3)_F$
symmetry as predicted by large $N_c$ considerations. 
In the case of 
$\mathcal{R}_{V/S1}$, 
moreover,
the difference is quite substantial, 
with an almost certain preference for the 
SU$(3)_F$-symmetric CQM value. 
Lattice QCD results for magnetic moments seem to suggest that large 
$N_c$
predictions are more robust in the 
$\text{SU}(3)_L \times \text{SU}(3)_R$ 
chiral limit, 
rather than the 
SU$(3)_F$ limit.
This finding should be contrasted with baryon mass relations in the large $N_c$ limit, 
which seem, 
however, 
to show the predicted improvement with 
SU$(3)_F$ symmetry%
~\cite{Jenkins:2009wv}.

\subsection{Mixing of $\Sigma^0$ and $\Lambda$ Baryons}
\label{s:siglam}

With the inclusion of a magnetic field, 
the two 
$I_3 = 0$
octet baryon states, 
$\Sigma^0$
and
$\Lambda$, 
mix through the isovector component of the quark charge matrix. 
This system hence requires a coupled-channels analysis, 
for which one forms the matrix of correlation functions
\begin{equation}
\mathcal{G}^{(s)}(t,n_\Phi)
=
\def\arraystretch{1.5}
\begin{pmatrix}
G_{\Sigma \Sigma}^{(s)}(t,n_\Phi)
&
G_{\Sigma \Lambda}^{(s)}(t,n_\Phi)
\\
G_{\Lambda \Sigma}^{(s)}(t,n_\Phi)
&
G_{\Lambda \Lambda}^{(s)}(t,n_\Phi)
\end{pmatrix}
,\end{equation}
where 
$s = \pm \frac{1}{2}$ 
is the spin projection along the magnetic field. 
The off-diagonal correlation functions in this system are only available in 
Ref.~\cite{Parreno:2016fwu}
on the 
SU$(3)_F$-symmetric ensemble.

\begin{figure} 
  \centering
  \sidecaption
  \includegraphics[width=9cm,clip]{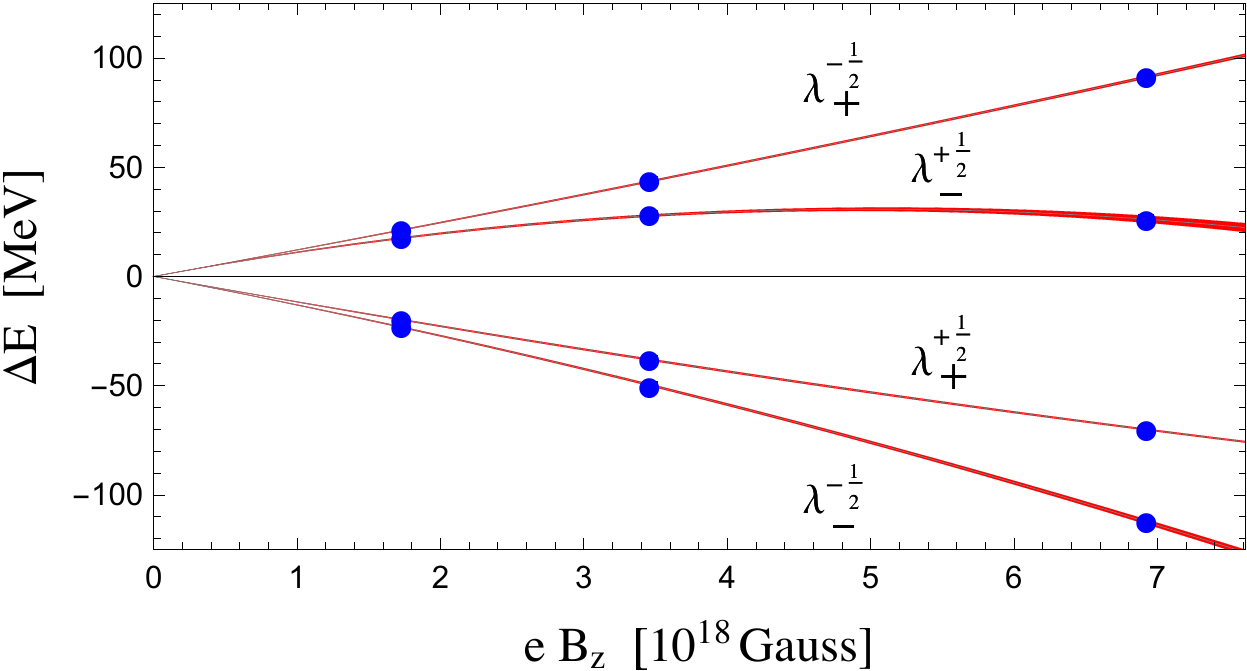}
  \caption{Energy eigenvalues 
  $E_{\lambda_\pm^{(s)}}$ 
  determined from a coupled-channels analysis in the $\Sigma^0$-$\Lambda$ system at 
  $m_\pi = 800 \, \texttt{MeV}$
  are plotted as a function of the magnetic field.
  Fits to the magnetic field dependence 
  predicted by Eq.~\ref{eq:Elamsig} are also shown.
  Note that for each energy eigenvalue,  
  the difference
  $\Delta E \equiv E - M_B$ is defined to vanish in vanishing magnetic field. 
  }
  \label{fig-27b}
\end{figure}

Eigenstate energies are extracted from the principal correlation functions that solve the generalized eigenvalue problem%
~\cite{Luscher:1990ck} 
posed in this system. 
Denoting the $U$-spin eigenstates as 
$\lambda_\pm$, 
the calculated energies are plotted as a function of the magnetic field strength in 
Fig.~\ref{fig-27b}. 
These energies are to be compared with those predicted on the basis of the
$U$-spin 
symmetry of the calculation, 
namely
\begin{equation}
E_{\lambda_\pm}^{(s)}
=
M_B
\pm
\mu_n \frac{s\, e B_z}{M_B}
-
\frac{1}{2} 4 \pi
\left[
\beta_n
+
(1 \pm 1)
\frac{2}{\sqrt{3}}
\beta_{\Lambda \Sigma}
\right]
B_z^2
+
\mathcal{O}(B_z^3)
\label{eq:Elamsig}
,\end{equation}
where 
$\mu_n$
is the neutron magnetic moment, 
and
$\beta_n$
is the neutron magnetic polarizability
(technically it is the quark-connected part of 
$\beta_n$ 
given the limitations of the calculation). 
The quantity 
$\beta_{\Lambda \Sigma}$
is the transition polarizability in this system. 
It does not have quark-disconnected contributions, 
and the level ordering observed in Fig.~\ref{fig-27b} 
requires 
$\beta_{\Lambda \Sigma} < 0$. 
From fits, 
the value
$\beta_{\Lambda \Sigma} = - 1.82(06)(12)(02) \, \texttt{fm}^3$
is obtained, 
which is to be compared with 
$\beta_n = 3.48(12)(26)(04) \, \texttt{fm}^3$, 
where the uncertainties reflect statistics, systematics, and scale setting.%
\footnote{ 
The lattice spacing is determined from the quarkonium hyperfine splitting without light quark-mass extrapolation, 
producing the value
$a = 0.1453(16) \, \texttt{fm}$.
Accounting for the light quark-mass dependence will result in a significant change to polarizabilities due to their 
$a^3$
scaling, 
\emph{cf}.~the scale setting of 
Ref.~\cite{Chang:2015qxa}.
}
With 
SU$(3)_F$
breaking, 
the 
$\Sigma$-$\Lambda$
mass splitting will suppress mixing, 
and consequently make it challenging to isolate the transition moment and polarizability from lattice QCD data.

\section{Further Study}
\label{s:end}

To conclude, 
we suggest calculations and questions for further study motivated by Ref.~\cite{Parreno:2016fwu} and the overview given above. 
1).~Concerning the pion mass dependence:
scaling by the baryon's Compton wavelength seems to account for a majority of the quark-mass dependence of octet baryon magnetic moments. 
This suggests that extrapolating from 
$800 \, \texttt{MeV}$ 
down to the physical point might be easier than extrapolating from the chiral limit to the physical point. 
Is there a way to understand this quantitatively, 
especially for the potential applicability to other observables? 
2).~The calculations presented here are limited in their tests of a few relations. 
A more complete study seems warranted, and this necessitates computing 
the decuplet magnetic moments and decuplet-to-octet transitions moments. 
3).~The flavor-breaking trend in large $N_c$ relations remains curious, 
and its study would be bolstered by further 
SU$(3)_F$-symmetric computations of magnetic moments at lighter quark masses. 
4).~The quark model can be more thoroughly vetted against large $N_c$ relations in computations with 
$N_c = 5, \cdots$. 
Nonetheless it appears that baryon magnetic moments, 
while seminal in the historical development of QCD, 
continue to provide interesting questions about non-perturbative physics.

\subsection*{Acknowledgements}

\footnotesize

Calculations were performed
using computational resources provided by the
Extreme Science and Engineering Discovery Environment
(XSEDE), 
which is supported by U.S.~National Science
Foundation grant number OCI-1053575, 
and NERSC,
which is supported by U.S.~Department of Energy Grant
Number DE-AC02-05CH-11231. 
The PRACE Research
Infrastructure resources Curie based in France at the
Tr\`es Grand Centre de Calcul and MareNostrum-III based
in Spain at the Barcelona Supercomputing Center were also
used. Development work required for this project was
carried out on the Hyak High Performance Computing and
Data Ecosystem at the University of Washington, supported,
in part, by the U.S.~National Science Foundation
Major Research Instrumentation Award, 
Grant Number 0922770. 
Parts of the calculations used the
Chroma software suite~\cite{Edwards:2004sx}.

WD was supported in part
by the U.S.~Department of Energy Early Career Research
Award No.~DE-SC00-10495 and by Grant Number DE-SC00-11090.
EC was supported in part by the USQCD
SciDAC project, the U.S.~Department of Energy through
Grant Number DE-SC00-10337, and by U.S.~Department
of Energy Grant No.~DE-FG02-00ER41132. 
KO was
supported by the U.S.~Department of Energy through Grant
Number DE-FG02-04ER-41302 and by the 
U.S.~Department of Energy through Grant Number 
DE-AC05-06OR-23177, 
under which JSA operates the Thomas
Jefferson National Accelerator Facility. 
The work of
AP is partially supported by the Spanish Ministerio de Economia y Competitividad (MINECO) under the project MDM-2014-0369 of ICCUB, and, with additional European FEDER funds, 
under the contract FIS2014-54762-P.
MJS was supported in part by 
U.S.~Department of Energy Grants No.~DE-FG02-00ER-41132
and No. DE-SC00-10337. 
BCT was supported in part by the U.S.~National Science Foundation, 
under Grant
No. PHY15-15738.


\bibliography{lattice2017}


\end{document}